\documentclass[
a4paper,
showkeys,
floatfix,
aps,
prl,
longbibliography,
superscriptaddress,
reprint,
]{revtex4-2}

\usepackage{graphics,graphicx}
\usepackage{amsmath,amssymb}

\usepackage{graphics,graphicx}
\usepackage{dcolumn,bm}
\usepackage{psfrag}
\usepackage{xstring}
\usepackage{color}
\usepackage[colorlinks=true,
linkcolor=blue,
citecolor=blue,
urlcolor=blue]{hyperref}
\usepackage{url}
\usepackage{float}
\usepackage[utf8]{inputenc}
\usepackage{placeins}

\newcommand{\srm}
{\affiliation{Department of Physics, SRM University - AP, Amaravati,
 Andhra Pradesh - 522240, India}}

\begin{document}

\title{Large earthquakes follow highly unequal ones}

\author{Sudip Sarkar}

\email{sudip\_sarkar@srmap.edu.in}
\srm
\author{Soumyajyoti Biswas}

\email{soumyajyoti.b@srmap.edu.in}
\srm

\begin{abstract}
It was conjectured for a long time that the tectonic plates are in a self-organised state of criticality and that the Gutenberg-Richter law is a manifestation of that. It was recently shown that for a system near criticality, the inequality of their responses due to external driving would sharply rise and show universal behavior that could indicate the proximity of the system to a critical point. As a result, measures such as the Gini and Kolkata indices that quantify inequality can also serve as indicators of imminent criticality and those of diverging (system-spanning) responses. In the context of earthquakes, such a large response would correspond to events of high magnitudes. In this work, we show with numerical simulations and seismic data analysis that large earthquake events have a tendency to follow events that are highly unequal, similar to the case of a system near a critical point. Even though this is not a proof of tectonic plate systems being near-critical, a continuous monitoring of the inequality indices of the earthquake time series could be an useful tool for hazard estimates. We have applied this framework to models of earthquakes as well as to the earthquake time series from various seismically active regions, such as North America, Southern Japan, parts of Southeast Asia and Indonesia. The findings indicate that the SOC picture of tectonic plates is consistent with the increase in size inequality of earthquakes, even though this cannot be treated as a rigorous proof.

\end{abstract}

\maketitle
\section{Introduction}

When tectonic plates move (at a very slow rate), they undergo a stick-slip motion that causes earthquakes when the stored potential energy is released due to a sudden slip event. Although much is known through more than a century of studies related to the motions and subsequent energy released in tectonic plates, its understanding still faces severe difficulties, particularly due to the inaccessibility of the regions of origin (often tens to hundreds of kilometers underground) \cite{carlson1994}. However, the phenomenon of earthquakes has attracted researchers from various disciplines, not only due to their devastating consequences but also due to their remarkable statistical regularities in time series \cite{kawamura2012}. In particular, earthquake magnitudes are known to follow the Gutenberg-Richter (GR) law \cite{gr1944}, $\log_{10}N=a-bM$, which states that the number of earthquakes ($N$) of magnitude $M$ is exponentially related to $M$ ($a$ and $b$ are constants). It is also observed that the rate of aftershocks decays with time ($t$) in a power law $n(t)=K/(t+c)^p$, known as the Omori-Utsu law \cite{omori,utsu}. There are other empirical laws about earthquake dynamics as well, which are verified using data analysis across various regions in the world. While a complete understanding of the mechanism behind earthquakes is still lacking, the empirical observations have been useful in gaining insights into the earthquake dynamics, improving the perspective to model the phenomenon and aiming for a forecasting possibility. The possible pathways for such forecasting include monitoring the changes in some statistical features of earthquake time series data prior to a large event, such as the widely observed lowering of the GR law exponent value \cite{scholz, kawamura, nanjo, takahiro}; elevated event and energy release rates \cite{schmit}; and unusual behavior in other complexity measures (see e.g., \cite{rundle1,rundle2,varo}). 

In this work, we analyse the correlation between the energy released in an earthquake and the inequality of the energies that were released by the events immediately prior to that earthquake. We show that measures of inequality, such as the Gini and Kolkata indices, estimated for a series of earthquake events within a time window, show universal features and precursory signals before an upcoming large event. It should also be mentioned that criticality is not the only mechanism that can explain such observations (lowering of the power law exponent and increase of $g,k$ indices). Alternatively, there are different models of earthquakes and real earthquake catalogs for several seismically active regions \cite{Ogata1988, Sornette2003, Helmstetter2003, seif2019}. 

An analytical explanation within reasonable assumptions could also be made. We further estimate the vicinity of the critical point, in terms of inequality indices and GR law exponent, explored by the system during its SOC-like dynamics. Such estimates of proximity to critical points unify the observation of lowered GR exponent (see e.g., \cite{nanjo}) and elevated event and energy release rates \cite{schmit} with increased inequality in energy released observed prior to large earthquake events in a compatible SOC-like framework. This work primarily focuses on the correlation between earthquake inequality and subsequent large events in the hope that such correlations could eventually be exploited for hazard assessments.

\section{Models and Methods}
 We first look at some of the models of earthquakes to check the inequalities of the avalanche series \cite{ofc}. Application of this method for such models has a wider consequence, given the broad domain of relevance for avalanche phenomena outside earthquake dynamics \cite{manna, first}.
Here we first describe the models used for this study and then move on to the inequality indices that are used.

\subsection{Sandpile like 2D SOC model of earthquake}

In this work we adopt the model introduced in ref. \cite{petrelis}. Each site is initialised with stress values taken randomly between ($0, S_c$). Then stress at each site increases at a constant rate $v_0$. When a site reaches a preassigned threshold $S_c$, an earthquake event is triggered. In defining the affected region, all clusters (identified through nearest neighbour active sites) consisting of sites having stress above $S_c-D_1$ are identified ($D_1$ is constant). For all sites within the cluster that now contains the triggering site, the stress value is reduced to $S_c-D_s$, where $D_s=D_2+D_3\eta_{ij}$, where $D_2$ and $D_3$ and $\eta_{ij}$ are uncorrelated random variables uniform between ($-1,1$) if the avalanche size (sites in the triggering cluster) is smaller than 5 and are Gaussian correlated random numbers between ($-3,3$) with a correlation length $D_d$ for avalanches larger than 5 \cite{lucas}. If $n$ is the number of sites in the triggering cluster, then the magnitude of the event is defined as $m=\frac{3}{2}\log n$. For GR law, the requirement is to set $D_3 \ll D_1 \ll D_2$. Here we have taken $v_0=1$, $S_c=10$, $D_3=0.1$, $D_1=5$, $D_2=10$ and system size $L=400$ (or $N=400\times 400$). A larger correlation would result in larger avalanches on average.

\subsection{Train model of earthquake}

We also simulate the train model, which is a simplification of the Burridge-Knopoff model \cite{bkmodel}, where a discrete set of massive blocks slides over a frictional surface that is also taken as a discrete set of fixed blocks \cite{vieira,biswas2013,ew}. The upper chain of blocks is connected through linear springs. When one end of the train is pulled, the blocks can move if the forces on it transferred through the elastic springs are sufficient to overcome the pinning force (friction) with the lower fixed block. The pinning force is drawn from a uniform distribution in ($0, 2$). The blocks are not allowed to move in the opposite direction. Then the number of blocks ($N$) moving as a result of the movement on the block at the end by one lattice space is defined as an ``avalanche". 

\subsection{ Inequality indices used}

\begin{figure}[t]
    \centering
    \includegraphics[width=\columnwidth]{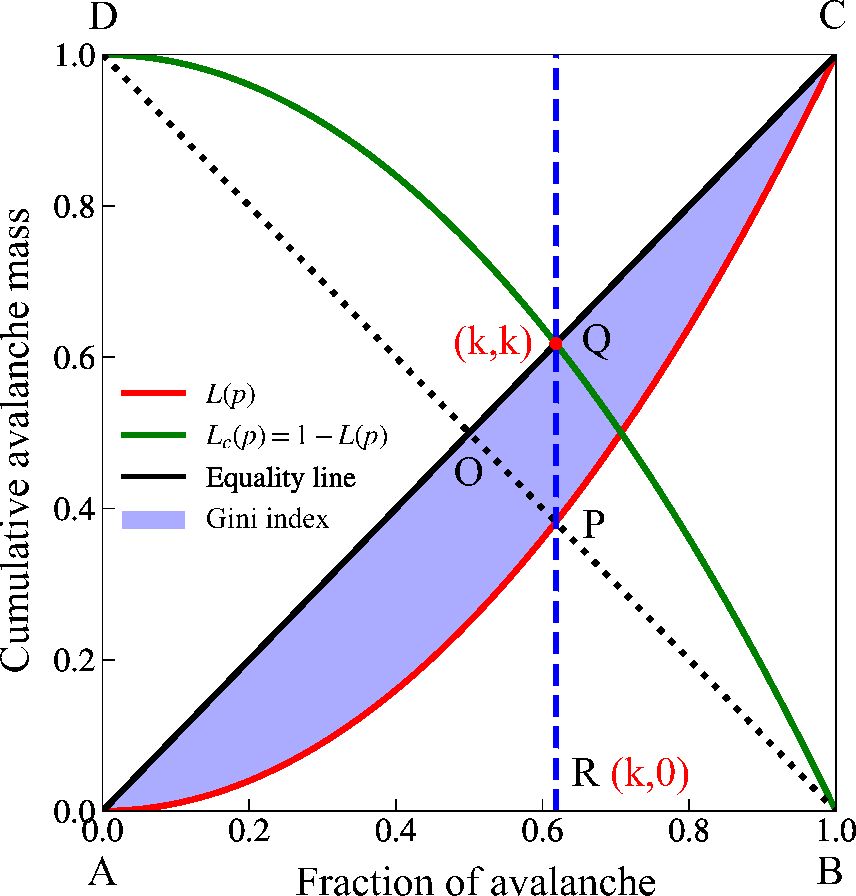}
    \caption{ A schematic diagram of the Lorenz function $L(p)$, equality line and the area representing the Gini index are shown. The Kolkata index is also shown as the intersection point between the Lorenz curve and the off-diagonal line.}
    \label{fig:lorenz}
\end{figure}

In quantifying the inequality of the earthquake events, in simulations and in the data, we use inequality indices that are traditionally used to quantify inequality of wealth distribution in a population, viz., the Gini \cite{gini} and Kolkata \cite{kolkata} indices. These indices can be measured from the Lorenz curve $L(p)$, which is defined as the cumulative fraction of energy released by the smallest $p$ fraction of the events in a sequence (Fig.~\ref{fig:lorenz}). Clearly, $L(0)=0$ and $L(1)=1$. If all events were of exactly equal size, then the Lorenz curve would be a diagonal line from (0, 0) to (1, 1). Any departure from the diagonal line is, therefore, a measure of inequality of the events in the chosen sequence. The Gini index is defined as the area between the diagonal line and the Lorenz curve, normalised by the area under the diagonal line (1/2). Formally, $g=1-2\int\limits_0^1L(p)dp$. The value of $g$ is always between 0 and 1, representing complete equality and complete inequality, respectively. The Kolkata index is defined as the largest $1-k$ fraction of the events that accounts for the $k$ fraction of the total energy released in the sequence. Formally, it can be obtained by solving $1-k=L(k)$. It is a generalisation of Pareto's 80-20 law \cite{pareto} in wealth (and many other) distributions, which states that $80\%$ of wealth is typically possessed by the $20\%$ of the richest individuals. These measures have been used recently in measuring inequality of the responses of systems approaching criticality and have been found to have universal features (see e.g., \cite{front}), particularly near a critical point. Here we use these for earthquake statistics (see also \cite{g_eq}), where the proximity to the critical point is perpetual as far as the SOC-like dynamics are concerned.

For a time series of events, we calculate $g, k$ cumulatively for all events in a sequence starting from $t_s$ to $t_e$ and assign those $g, k$ values to the event at $t_e + 1$, i.e., the subsequent event in the sequence. When a large event takes place, say, at $t=t^{\prime}$, we reset $t_s=t^{\prime}$ and keep on calculating $g,k$ cumulatively for the successive events and assign those values to the subsequent event as before. In this way, all events will have $g,k$ values assigned to them, calculated from a dynamic time window between the last large event up until the just preceding one. The code is publicly available in \cite{code}.

Note that the calculations of the inequality indices are also done using a sliding time window of fixed size. But the resulting correlation is weaker, especially for the earthquake catalog. The reason is that for a fixed window, even after a large event has passed, as long as it remains within the window size ($W$), the inequality indices ($g,k$) will be large. In the following, we compare the results from both the methods.

\section{results}
Here we report the inequality of earthquake events from the models and data. In earlier studies ~\cite{jordi}, it was seen that system-spanning events often happen when $g,k$ grow larger and cross each other (around $0.865$). Here we analyse the time series for earthquake models and catalog data.
\subsection{ Inequality of avalanches in the models }

\begin{figure}[t]
    \centering
    \includegraphics[width=\columnwidth]{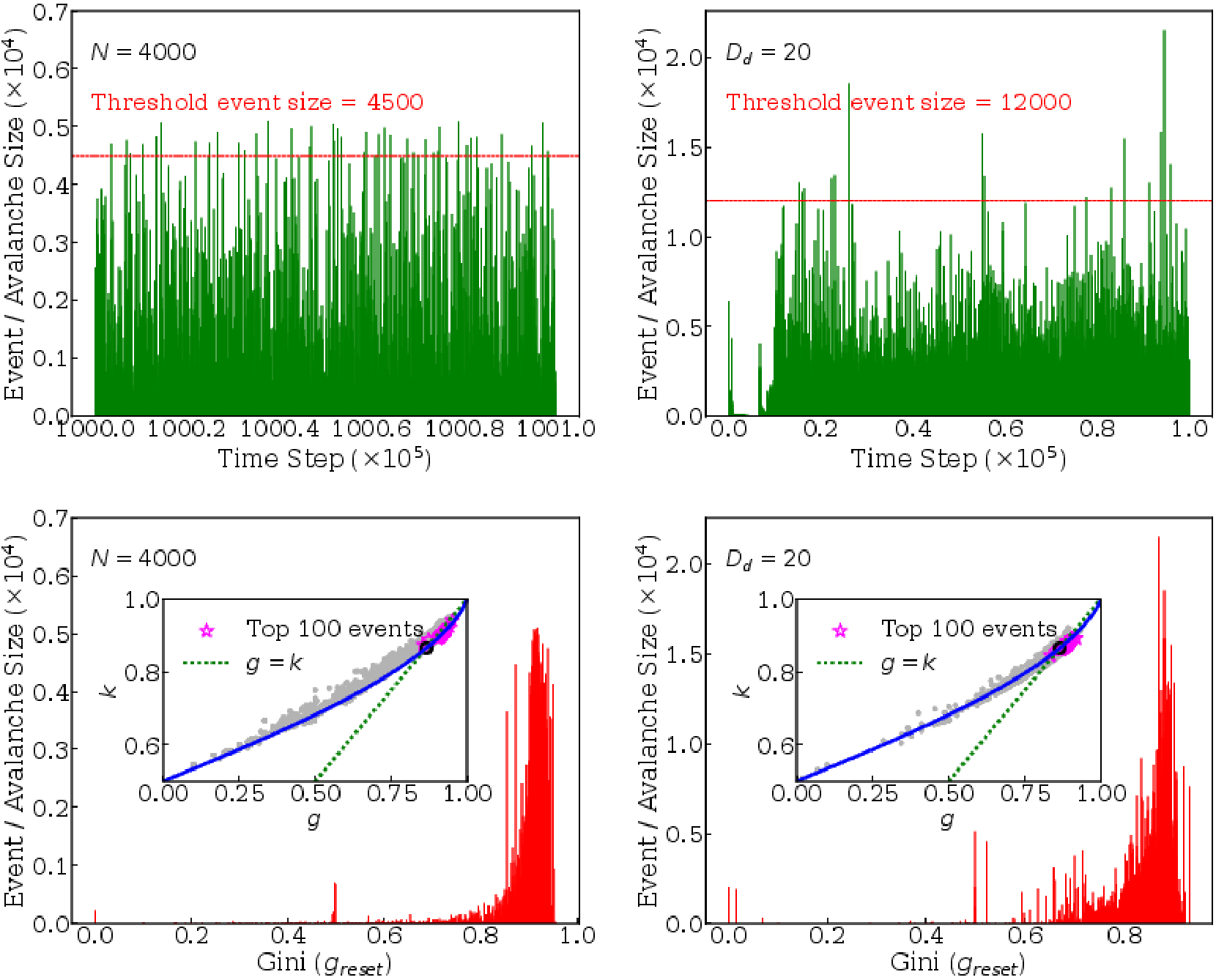}
    \caption{
    The top row figures show the time series of the avalanches seen in the 2D SOC and train models. The bottom row figures are the same time series but now sorted according to the $g$ values associated with each event. The $g$ (and $k$) value associated with an event is calculated from the time window just preceding that event, with the width dynamically determined by the reset after a large event (see text). A clear correspondence is noted between high $g$ values and large avalanches.
    }
    \label{fig:model_reset}
\end{figure}

Once the avalanche time series are obtained in the model simulations, we sort the events according to the associated $g$ values, as mentioned above. Fig.~\ref{fig:model_reset} shows the typical nature of such sorting. While there is a dependence (hence arbitrariness) on the choice of a ``large" event, we see a clear correlation between higher $g$ values and larger magnitudes/avalanche sizes. Similar correlation is expected in $k$. The insets show the $k$ vs $g$ plots with the largest 100 events highlighted. This shows that the large events happen generally above the $g=k\approx 0.865$ point.

\subsection{Inequality of energy released in earthquakes}

We then analyze real earthquake data from various parts of the world (Southern Japan, Southeast Asia, North America, and Indonesia). The coordinates used are as follows: Southern Japan (30.145°–41.278° N, 129.792°–142.119° E); Southeast Asia (20.633°–35.747° N, 67.588°–99.000° E); North America (12.383°–61.143° N, 145.547°–46.318° W); and Indonesia (7.537° S–5.703° N, 95.142°–107.051° E). The duration is from January 1975 to October 2025. The catalog used (USGS) can be found in \cite{usgs}.

\begin{figure}[t]
    \centering
    \includegraphics[width=\columnwidth]{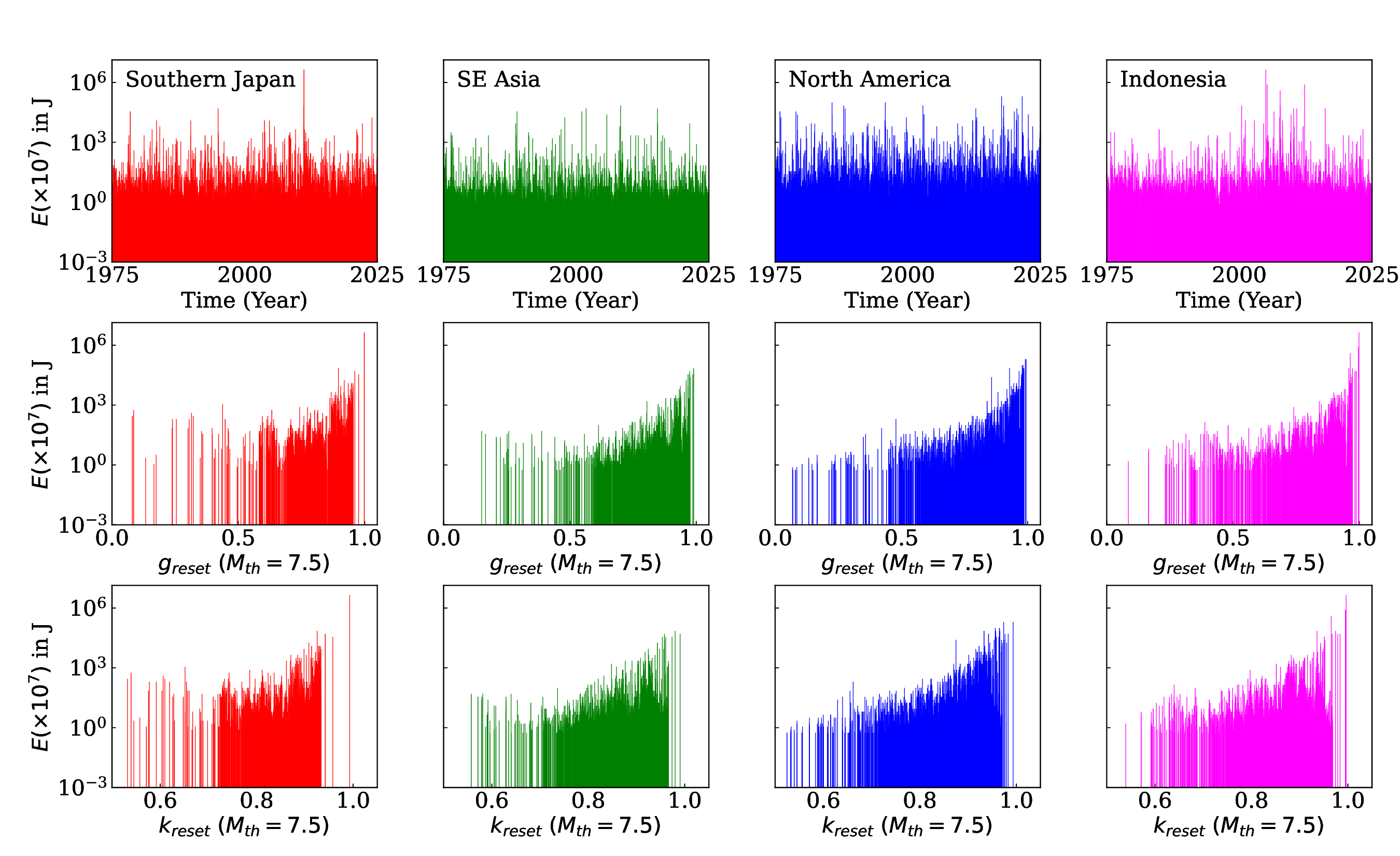}
\caption{ The top row figures show the earthquake event catalog from different earthquake-prone regions of the world in their original order of occurrence. The middle and bottom row figures show the same events when arranged in terms of their associated $g$ and $k$ values (indices), respectively, where the indices for any event are calculated from a dynamic time window defined as all events between the previous large event ($M\geq 7.5$) and the current event (excluding both ends). A clear correlation between higher inequality and larger earthquake magnitude is seen for each of the seismically active regions.}
    \label{reset_order}
    \end{figure}

The lower cut-off used in magnitude is $4.5$. Then the magnitudes are converted to energy \cite{moment} using,
\begin{equation}
\log_{10} E = 1.5M + C
\end{equation}

\noindent where the prefactor $C$ does not influence the inequality measures. The completeness of the resulting catalog was checked by noting that the size distribution follows GR law. Note that inequalities in the magnitudes were reported very recently in \cite{g_eq}. In Fig.~\ref{reset_order}, the unsorted (actual order) and sorted (according to $g$ values) series are shown for earthquake sequences in the different regions. Once again, a clear trend of accumulations of larger events towards higher $g$ values is observed. A similar trend is observed when the events are sorted according to their $k$ values.

\begin{figure}[tbh]
    \centering
    \includegraphics[width=\columnwidth]{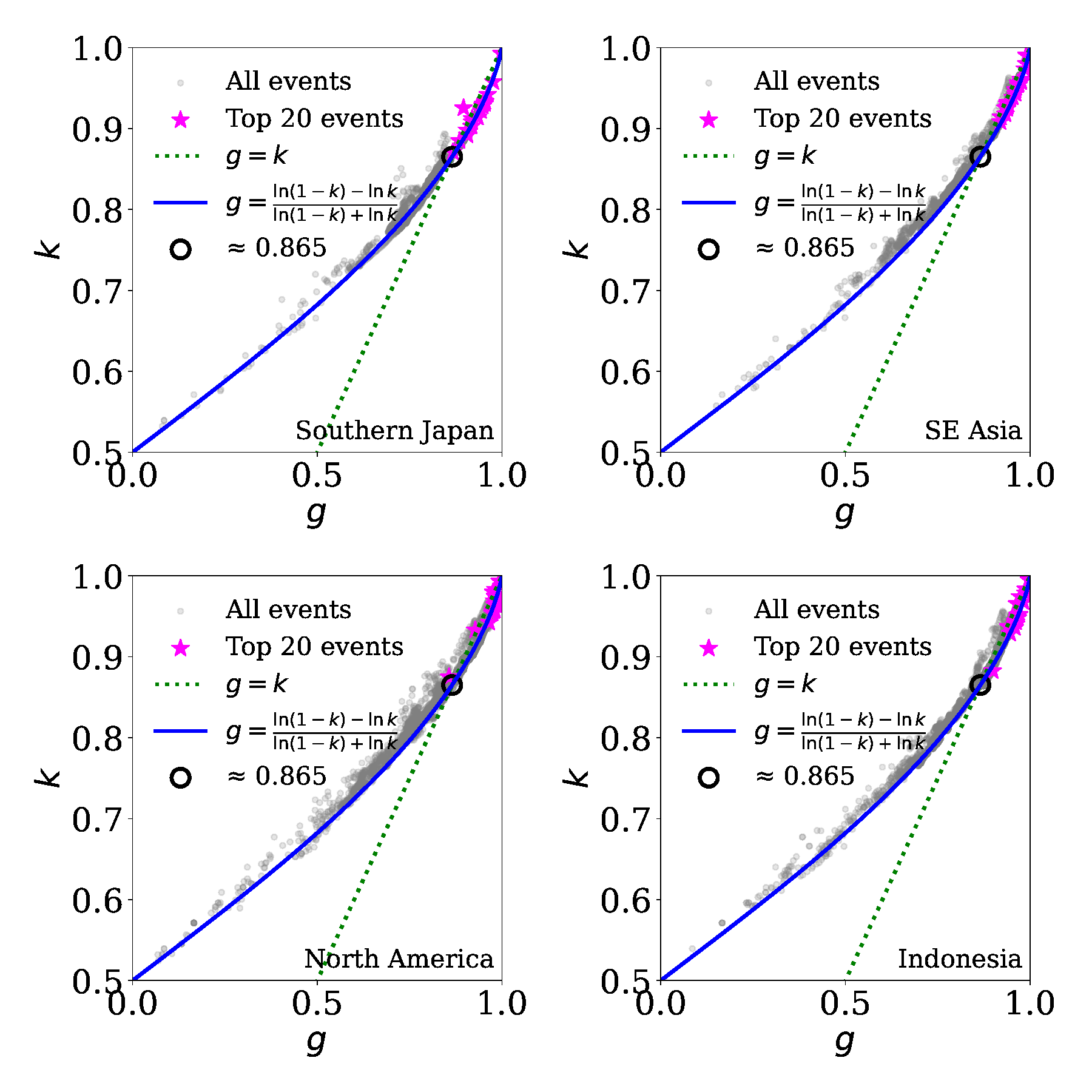}
\caption{ For the earthquake catalog data of different regions, the $g$ and $k$ values obtained from the dynamic reset method (see text) are plotted, and the largest 20 events are highlighted. All of the largest $20$ events happen around $g\approx k$ and for higher values of the inequality indices.}
    \label{fig_reset}
    \end{figure}

In Fig.~\ref{fig_reset}, we plot $k$ vs. $g$ for each event (with larger ones highlighted). All points fall around a common characteristic shape $g=\frac{\ln(1-k)-\ln(k)}{\ln(1-k)+\ln(k)}$ (as given in eq.~\ref{gk_relation}), which is independent of the exponent of the GR law \cite{front}. The top $20$ largest events in each region fall above the intersection of the $g$ vs $k$ curve and the straight line $g=k$, indicating larger values of the inequality indices, and particularly their crossing can signal imminent large events.

\subsection{Correlation between inequality indices and earthquake (avalanche) size}

It is possible to show these behaviors of the inequality indices starting from the GR law under certain approximations. It is also possible to relate the elevated inequality of the sizes of events prior to a large one with the lowering of the GR law exponent ($\beta$). It is known that for power law size-distributed events, say following $P(S)\sim S^{-\beta}$ where, $S$ is the avalanche (event) size. Of course, there is always a physical lower and upper cut-off. If $r_0$-th event is the largest one, then if all events are arranged in ascending order, the $r$-th event size would follow $S_r \sim (r_0-r)^{-n}$, where the diverging/system-spanning event happens at $r=r_0$, with $n=\frac{1}{\beta-1}$ (see e.g., \cite{eswar}. For a system driven towards a failure/transition, the natural order is often close to the ascending order. For the systems with near-stationary sequences, precursory events show a growing rate of energy release prior to large events (see e.g., \cite{jordi}). Thus, a reasonable assumption is the (nearly) ascending order of events prior to a large event (see e.g., \cite{schmit}). It should be noted, however, that the assumption is better suited for systems leading up to a catastrophic failure, rather than those generating a stationary sequence. For a segment $r=0$ to $r=xr_0$ (assuming that the a chosen sequence of events fall within that segment), the Lorenz function corresponding to the avalanches would be

\begin{equation}
L(p,x,n)=\frac{\int\limits_{0}^{pxr_0}(r_0-r)^{-n}dr}{\int\limits_{0}^{xr_0}(r_0-r)^{-n}dr},
\label{lorenz_int}
\end{equation}

\noindent which leads to  
\begin{equation}
L(p,x,n)=\frac{1-\left(1-px\right)^{1-n}}{1-(1-x)^{1-n}}.
\label{lorenz}
\end{equation}

\noindent One can then evaluate the Gini index in the closed form \cite{prl} as,

\begin{equation}
\begin{split}
g(x,n)=&\,1-2\int_0^1L(p,x,n)\,dp \\
=&\,1-\frac{2}{1-(1-x)^{1-n}}
\left[
1+\frac{(1-x)^{2-n}-1}{(2-n)x}
\right],
\end{split}
\label{g_eq}
\end{equation}

\noindent where $x=\frac{r}{r_0}$. The critical point (for a finite system, the point of largest avalanche) is $x=1$. At that point,
\begin{equation}
    L_f(p,n)=L(p,n,x=1)=1-(1-p)^{1-n}.
\end{equation}

\noindent Note that the Lorenz function has a scaling form
\begin{equation}
    L(p,n,x)=\frac{L_f(px,n)}{L_f(x,n)}.
\end{equation}

\noindent The Kolkata index follows $1-k=L(k,x,n)$, giving the general relation between $g(x,n)$ and $k(x,n)$ as

\begin{equation}
    g(x,n)=1-\frac{2(1-k(x,n))}{1-(1-k(x,n)x)^{1-n}}[1+ \frac{(1-x)^{2-n}-1}{(2-n)^x}].
    \label{general_g}
\end{equation}

\noindent For $x\rightarrow 1$ Eq. (\ref{lorenz}) becomes

\begin{equation}
    L(p,n)=1-(1-p)^{1-n}
\end{equation}

\noindent as before $n=\frac{1}{\beta-1}$. This gives

\begin{equation}
    g(n)=1-2\int_{0}^{1} L(p,n)dp =\frac{n}{2-n}.
\end{equation}

\noindent Also, by the definition of the Lorenz function (Eq. (~\ref{general_g})) gives $1-n=\frac{lnk}{ln(1-k)}$. It then gives the simpler (n-independent) form 

\begin{equation}
    g=\frac{ln(1-k)-lnk}{ln(1-k)+lnk}.
    \label{gk_relation}
\end{equation}

\noindent The reason that most points (both in simulations and real data) for $x<1$ nevertheless follow the $x\rightarrow 1$ limit of the relation is that the $x$ and $n$ dependencies are weak in Eq. (~\ref{general_g}).

To show the positive correlation between avalanche size and the Gini index value of the preceding avalanches, we can show it $\frac{dg}{dx}>0$ from Eq. (~\ref{g_eq}). Also, writing $S(x)=r_0^{-n}(1-x)^{-n}$, we see $\frac{dS}{dx}=nr_0^{-n}(1-x)^{-n-1}>0$. Therefore, 
 
\begin{equation}
\frac{dg}{dS}=\frac{dg}{dx}\frac{dx}{dS}>0,
\end{equation}

\noindent for all $n>0$, except at $n=1$ and $n=2$, where the derivatives don't exist. 
This positive correlation between an avalanche size and the inequality in preceding avalanches is reflected in the earthquake catalog data (Figs.~\ref{reset_order},~\ref{fig_reset}) and in simulations of SOC-like models (Fig.~\ref{fig:model_reset}). 

\begin{figure}[t]
    \centering
     \includegraphics[width=\columnwidth]{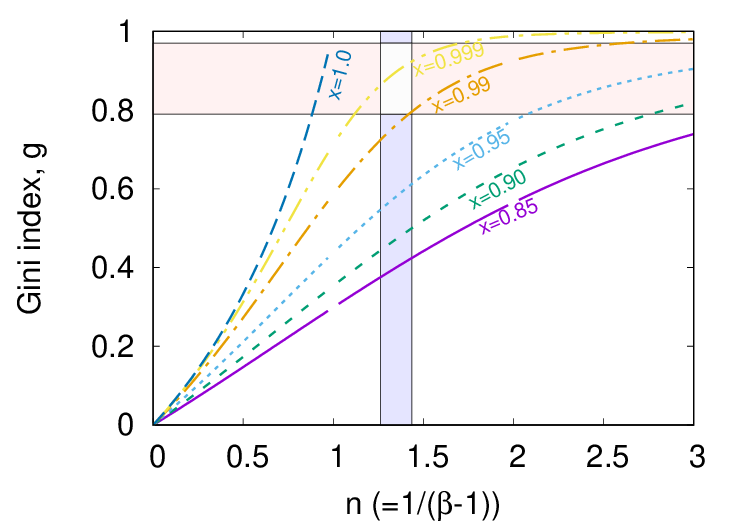}
\caption{ The variations of the Gini index ($g$) are shown with $n$ (hence, the event size distribution exponent $\beta$) for various values of the parameter $x$ that denotes proximity to a critical point ($x=1$ is the critical state; see text for details). The shaded horizontal rectangle denotes the range of $g$ values observed in the dynamic reset (see Fig. \ref{fig_reset}) method: $\langle g\rangle\pm \sigma_g$, where $\langle g\rangle=0.88$ is the average and $\sigma_g=0.09$ is the standard deviation. The shaded vertical rectangle denotes the range of $n$ (calculated from the range of $\beta$ obtained in the GR law fit). The overlapping shaded area is an indicator of where the system might be residing during the course of the dynamics, and it is always very close to the critical point, as is conjectured in an SOC picture of tectonic plates.}
    \label{n-g-x}
    \end{figure}

The dimensionless parameter $x$ is an indicator of proximity to the critical point (for example, $\frac{T}{T_c}$ in magnetic systems, $\frac{\sigma}{\sigma_c}$ in fracture models and so on). A similar interpretation applies for SOC systems even though such parameters are not explicitly tuned in those systems (e.g., a normalised height $\frac{h}{h_c}$ in sandpiles). In Fig. \ref{n-g-x}, we plot the variation of $g$ as a function of $n$ (hence, $\beta$) for different values of $x$. Within the plot, we indicate the $g$ values obtained through the reset method mentioned above. Note that an increase in the $g$ value would correspond to either an increase in $n$ (implying a lowering of the $\beta$ value) or a closer proximity to the critical point (increase in $x$), or both. Therefore, the lowering of the $\beta$-value, observed for many large earthquakes \cite{nanjo}, is consistent with an increase in the inequality of event sizes prior to a large event. Even though a range in the $n$ values is indicated from the error margin of fitting the GR law from the data, the changes in the exponent could be more than the range, but it is generally difficult to estimate for few events.

The intersection area of the two rectangular regions in Fig. \ref{n-g-x}, indicating the observed $g$ (within one standard deviation) and $n$ (within fitting error bars) values, depicts the area where the tectonic plate system is likely to reside over the course of its dynamics. As expected in the case of an attractive fixed point at criticality, a characteristic of SOC dynamics, the system is always very close to the critical point $x=1$. The resulting scale-free nature of response (avalanche) statistics creates the characteristic variations of the inequality indices that can eventually be used for estimating the hazard of imminent large events, as almost all large events follow the crossing of $g$ and $k$.

\subsection{Power law exponent study with safe and warning zone events}

\begin{figure}[t]
    \centering
    \includegraphics[width=\columnwidth]{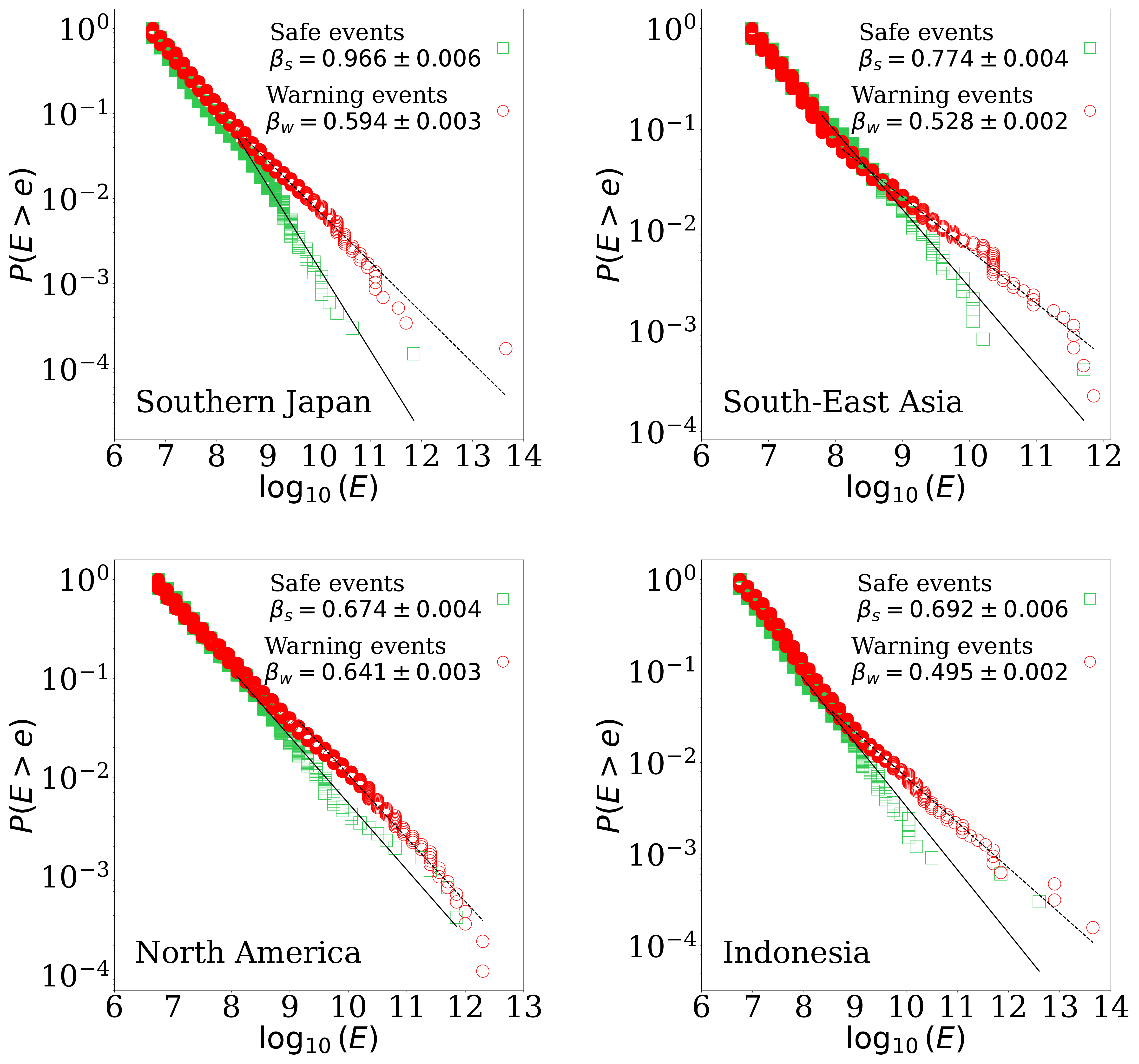}
    \caption{Cumulative frequency-energy distributions are plotted separately for the total events belonging to the safe and warning zones; the exponent value corresponding to the safe events, denoted by $\beta_s$, is found to be smaller than that for the warning events, denoted by $\beta_w$. This observation is consistent with the well-established precursor behaviour in earthquake prediction studies, where the exponent value tends to decrease prior to the occurrence of a large earthquake event.}
    
    \label{fig:s_w_exp}
\end{figure}

From the $g,k$ time-series plots, we can see that before any large earthquake (event), the $g$ index crosses the $k$ index, which is also seen in other fracture experiments \cite{jordi} and models. By taking this as a precursory indicator, if the events are classified into two categories based on the conditions $g \leq k$ (``safe'') and $ g > k$ (``warning''), and the cumulative frequency-energy distributions are plotted separately for the total events belonging to the safe and warning zones, then the exponent value corresponding to the safe events, denoted by $\beta_s$, is found to be smaller than that for the warning events, denoted by $\beta_w$ (see Fig.~\ref{fig:s_w_exp}). This observation is consistent with the well-established precursor behavior in earthquake prediction studies, where the exponent value tends to decrease prior to the occurrence of a large earthquake event.  We also examine this behavior for each reset window in all the regions. 
However, the number of events within the safe and warning windows is too small to construct reliable cumulative frequency-energy distributions and fit them with a power law, resulting in large fitting errors.

\subsection{Precursory signals using inequality indices and its statistical significance}

We have observed that $g,k$ rises before significant events. We now try to apply this to forecasting analysis. The ``warning signal" can be set in two different ways. First, we can think about the case where $g>k$ (as in fracture); second, we can establish various $g,k$ thresholds when the $g,k$ values exceed the threshold. For $g,k$ crossing, almost all ``large" events seem to be above $g\approx k \approx 0.865$ (see Fig.~\ref{fig_reset}). The events before crossing and after crossing show different power law exponent values (see Fig.~\ref{fig:s_w_exp}). Setting thresholds for $g,k$ and event size helps systematic analysis. We use a receiver operating characteristic (ROC) to quantify the predictive capability of the inequality indices in models and real data. ROC analysis is a common technique used to evaluate the performance of a binary classification model. We defined a threshold earthquake magnitude ($M_{thr}$) or avalanche size ($S_{thr}$) to classify an event as ``large". Also, the threshold values of $g$ and $k$ are then varied systematically as $g_{thr}$ and $k_{thr}$. If any event has magnitude (avalanche size) $M\geq M_{thr}$ then it is classified as ``Actual large event", and if the magnitude (avalanche size) $M<M_{thr}$, it is classified as ``Actual small event". To show the predictive behavior by using the inequality indices, we define ``Predicted large event" as $g\geq g_{thr}$ and ``Predicted small event" as $g<g_{thr}$. A large event is correctly predicted as large, i.e., a true positive (TP), if \(M \geq M_{\mathrm{thr}}\) and \(g \geq g_{\mathrm{thr}}\). A large event is incorrectly predicted as small, i.e., a false negative (FN), if \(M \geq M_{\mathrm{thr}}\) but \(g < g_{\mathrm{thr}}\). A small event is correctly identified as small, i.e., a true negative (TN), if \(M < M_{\mathrm{thr}}\) and \(g < g_{\mathrm{thr}}\). Finally, a small event is incorrectly identified as large, i.e., a false positive (FP), if \(M < M_{\mathrm{thr}}\) but \(g \geq g_{\mathrm{thr}}\). We can also do the same with Kolkata index. 

The ROC curve is obtained by plotting TPR versus FPR, and the predictive performance is measured using the area under the curve (AUC). An AUC value of 0.5 corresponds to random prediction, while larger AUC values show better predictive capability of the inequality indices in identifying large earthquake events or avalanches. TPR and FPR are defined as 

\begin{equation*}
R\,= TPR\, =\, \frac{TP}{TP+FN}
\end{equation*} and 

\begin{equation*}
FPR\, =\, \frac{FP}{FP+TN}.
\end{equation*}

\noindent We also use the $F_1$ score as an additional criterion to select the optimal threshold for the ROC analysis. The \(F_1\) score is defined as

\begin{equation*}
F_1 = \frac{2PR}{P+R},
\end{equation*}

\noindent where, the precision (P) is given by

\begin{equation*}
P = \frac{TP}{TP+FP}.
\end{equation*}

\noindent At each point on the ROC curve, the corresponding $F_1$ score is calculated. We then identify the maximum $F_1$ score for that ROC curve. This procedure is repeated over all event-size thresholds. The threshold whose ROC curve gives the largest $F_1$ score is selected as the ``best $F_1$ threshold". The AUC is then calculated from the complete ROC curve corresponding to this selected threshold. Therefore, the $F_1$ score is used to select the threshold, while the AUC is still calculated from the ROC curve.

\begin{figure}[t]
    \centering
    \includegraphics[width=\columnwidth]{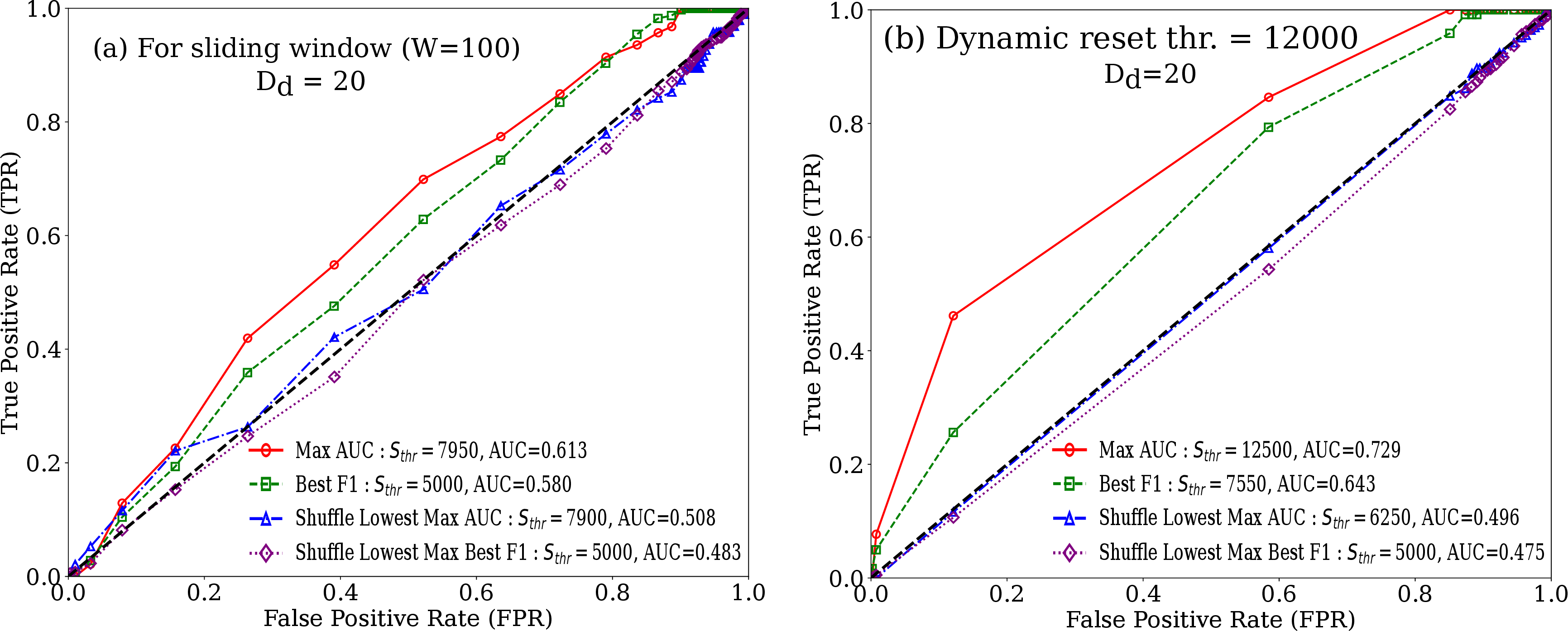}
    \caption{ROC analysis for the 2D SOC model with a correlation parameter $D_d=20$ using the inequality indices determined by (a) a sliding window of $W=100$ and (b) dynamic reset. Compared to the sliding-window method, dynamic reset gives a higher AUC value. We also check the ROC for the train model, but the AUC value approaches random prediction due to the large number of false positives.}
    
    \label{fig:2d_roc}
\end{figure}

To test the significance of the observed correlation (as in Figs.~\ref{fig:model_reset} and ~\ref{reset_order}), we randomly reassign the calculated $(g,k)$ pairs to the events for both the models and real datasets. In this shuffling procedure, $g$ and $k$ are kept together as a pair and are not shuffled independently. This preserves the internal relation between them, while destroying their temporal association with the corresponding event sizes. As a result, the correlation between large $g,k$ values and large event sizes becomes weaker. This randomization is repeated 1000 times. The ROC plots for shuffled data shown in Figs.~\ref{fig:2d_roc} and \ref{fig:roc_all}, made with one of the shuffled data among 1000 shuffled datasets for which we get the lowest maximum AUC and best F1-based AUC.

For North America and Southeast Asia, the AUC approaches values close to random prediction. One possible reason is the use of very large geographical regions that combine multiple tectonic environments, fault systems, and seismic regimes into a single dataset. Such spatial heterogeneity can reduce localized precursory patterns and also reduce the predictive skill of statistical indicators. Earthquake occurrence is strongly controlled by regional tectonic conditions, stress accumulation processes, fault interactions, and seismic clustering behavior, all of which can vary substantially across large areas. Consequently, combining multiple tectonic settings may average out region-specific precursor signatures and weaken forecasting performance  \cite{Zaccagnino_2023, wang_2025}. Catalog-related issues (catalog incompleteness) may also contribute to the observed behaviour \cite{mizrahi_2021}.

\begin{figure}[t]
    \centering
    \includegraphics[width=\columnwidth]{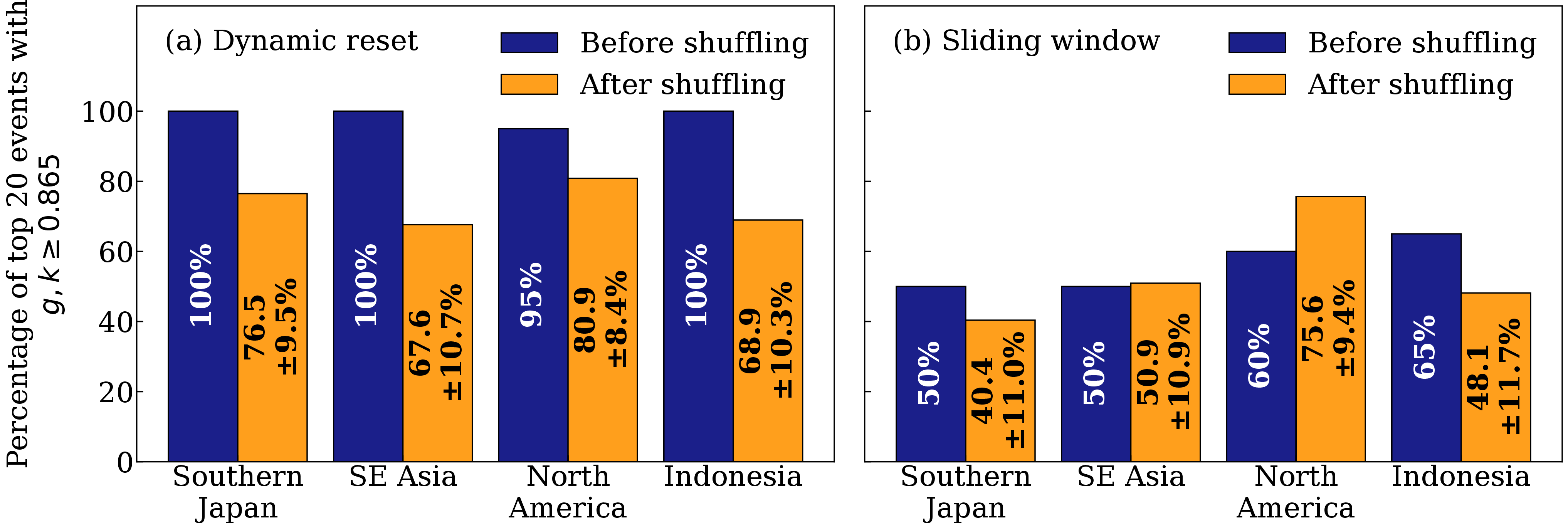}
    \caption{These figures show how many large events (percentage) out of the top 20 large events are coming above the intersection point $g\approx k\approx 0.865$ before we shuffle the $g,k$ values and after we shuffle the $g,k$ values (assigning them randomly to the events). For each region, we shuffle the inequality-index pairs $g,k$ 1000 times and then calculate the average number of events, out of the top 20 largest events, that lie above $g \approx k \approx 0.865$. This plot shows that before the shuffling, almost all the top 20 events are coming above the intersection point for the ``dynamic reset" method, but whenever we shuffle them, the number of top events counted above the intersection point decreases, which is consistent for all the regions. But when we are checking the same thing for the ``sliding window" method (with W=100), before shuffling, about $50\%-65\%$ events lie above the intersection point. Also, for two regions (Southeast Asia and North America), the percentage of top event counts increases after shuffling. }
    
    \label{fig:usgs_event_count}
\end{figure}

To examine the effect of shuffling, we count the number of events among the top 20 largest events that lie above the intersection point of the $g \approx k$ line and Eq.~\ref{gk_relation}. In the case of ``dynamic reset" (as shown in Figs.~\ref{fig_reset} and ~\ref{fig:usgs_event_count}(a)) before shuffling, most of the top 20 events ($95\%-100\%$), occur above $g \approx k \approx 0.865$. In contrast, after shuffling, the average number of top 20 events above this point decreases. Additionally, we verify this for the sliding window method. Fig.~\ref{fig:usgs_event_count}(b) illustrates that the percentage of significant events prior to shuffling (only $50\%-65\%$) is lower than that of the dynamic reset method. 

\begin{figure*}[!t]
    \centering
    \includegraphics[width=1.0\textwidth]{ROC_all.eps}
    \caption{For different earthquake-prone regions, we construct ROC curves and compute the corresponding AUC values using $g$ and $k$ inequality indices which are determined through (a) the dynamic reset method and (b) the sliding-window method with $W=100$.}
    \label{fig:roc_all}
\end{figure*}

\section{Discussion and Conclusion}

The inequalities of the earthquake energy released in models and data show a consistent trend of the larger events following a set of highly unequal ones. This is seen in the two models studied here, a 2D SOC model \cite{petrelis} and the train model of earthquake \cite{biswas2013} and also in the data taken from various earthquake-prone regions around the world. While earthquake dynamics have been cast in the form of SOC dynamics over many years, there are some persistent doubts as well. However, at least for the behavior of inequality indices, such dynamics seem to be consistent with the near-universal trend observed in the SOC models \cite{manna} (also see the discussions on Fig. \ref{n-g-x}). In the context of earthquakes, this also aligns with the earlier observations of the lowering of the size distribution exponent ($\beta$) prior to large events \cite{kawamura, nanjo}. A lowering of $\beta$ necessarily means an increase in the inequality of the events (see Fig. \ref{n-g-x}). Therefore, in that context, this current observation can provide an alternative to fitting the exponent values from a limited dataset, and just calculating their inequality could work in providing a potential hazard analysis. On the same note, the estimates of the GR law exponent (size distribution exponent) and the measured values of $g$ together suggest a very close proximity of the model to a critical point. While this is not a rigorous proof, it gives an indication of SOC-like dynamics in tectonic dynamics in a quantitative manner. On the other hand, a close proximity to the critical point tends to increase the inequality in response preceding events that are not necessarily always very large in comparison to the other events in a different part of the tectonic plate, since that would depend on the specific areas of the plate boundaries considered. The inequality indices considered here do not depend on the absolute sizes of the events but only depend on the functional form they follow. Therefore, we notice many small events for large $g,k$ values that could be large compared to the other local events but not large globally. However, for the most part large events do not occur for small $g,k$ values. The elevated $g,k$ values prior to a large event are also consistent with a significant increase in the event and energy release rates, which were observed for earthquakes \cite{schmit} and also more recently for compression failure of nanoporous materials \cite{jordi}.

In our analysis, the inequality indices are calculated from the sequence of event sizes (time step), not from the actual event times. Therefore, direct effects of temporal clustering are reduced. Also, shuffled catalogs, which preserve the event size distribution, do not show the correlation between inequality of event sizes and subsequent large events. This suggests the importance of event sequence and, hence, the correlation. However, this does not prove that the system is critical. Rather, the results are consistent with a critical-like picture. Similar caution has been suggested in studies showing that apparent temporal organization in earthquake catalogs can be affected by catalog incompleteness and stochastic triggering processes ~\cite{mizrahi_2021,davidsen2011,zhuang2023}. Thus, our results are best viewed as being consistent with a critical-like interpretation rather than as uniquely establishing it.

In the ``dynamic reset" procedure the sequence is restarted after a large event, and the inequality indices are then assigned to subsequent events. This choice is physically connected with the idea that a large event may partially release accumulated stress and initiate a new loading cycle. The ROC analysis (see Figs.~\ref{fig:2d_roc} and ~\ref{fig:roc_all}) indicates that both the ``dynamic reset" and ``sliding-window" methods possess predictive potentials for smaller geographical regions. However, the results of the shuffling test reveal that the dynamic reset method is more effective in capturing large top events than the sliding-window approach, as illustrated in Fig.~\ref{fig:usgs_event_count}. Consistently, the correlation between the occurrence of large events and elevated values of the inequality indices is stronger for the dynamic reset method. This suggests that resetting the analysis after each large event preserves information relevant to the subsequent buildup process. In contrast, the sliding-window approach may be more appropriate for systems that evolve toward a single catastrophic failure, such as rock fracture experiments, where the time series terminates at the final failure event.

Other physical processes may also produce similar patterns before large earthquakes. These include brittle–ductile coupling, variations in friction along the fault, afterslip-driven aftershock activity, and slow preparation before failure. For example, if aftershock activity follows the rate of afterslip, it can naturally produce Omori-like decay. Similarly, models with heterogeneous friction can reproduce both aftershock and foreshock patterns, including changes in the GR-law exponent value. In this sense, foreshocks may represent a real preparatory stage before the mainshock, rather than only a general critical buildup, as suggested by earlier studies \cite{avouac2004, mignan2014}. Therefore, further comparison with ETAS-like models \cite{Ogata1988, seif2019, petrillo2021} and more realistic physics-based simulations is needed.

In conclusion, we have studied earthquake energy size inequalities in models and real data from some of the seismically active areas around the world and found that there is a correlation between inequality and earthquake size (avalanche). This result is also compatible with a power law exponent $\beta$ or, more commonly, a decrease in the Gutenberg-Richter exponent value and can be used for precursory signals, which are more suitable when data analyses are confined to smaller regions.

\section*{Acknowledgement}

We thank Bikas K. Chakrabarti and Takahiro Hatano for useful discussions. The simulations have been performed in the HPCC Chandrama at SRM University-AP.

\section*{Data Availability statement}
The codes used in this study are freely available in Ref. \cite{code}

\FloatBarrier

\end{document}